\documentclass[twocolumn,showpacs,floatfix,superscriptaddress,amsmath,amssymb,prl]{revtex4}
\usepackage{mathrsfs}
\usepackage{txfonts}
\usepackage{amssymb}
\usepackage{graphicx}
\usepackage{hyperref}
\usepackage{ulem}
 \usepackage{overpic}
 \usepackage{psfrag}
 \newcommand{\be}{\begin{equation}}
\newcommand{\ee}{\end{equation}}

\begin{document}
\title{Bifurcation in ground-state fidelity for a one-dimensional spin model with competing two-spin and three-spin interactions}

\author{Hong-Lei Wang, Yan-Wei Dai, Bing-Quan Hu, and Huan-Qiang Zhou}
\affiliation{Centre for Modern Physics and Department of Physics,
Chongqing University, Chongqing 400044, The People's Republic of
China}

\begin{abstract}
A one-dimensional quantum spin model with the competing two-spin and
three-spin interactions is investigated  in the context of a tensor
network algorithm based on the infinite matrix product state
representation. The algorithm is an adaptation of Vidal's infinite
time-evolving block decimation algorithm to a translation-invariant
one-dimensional lattice spin system involving three-spin
interactions. The ground-state fidelity per lattice site is
computed, and its bifurcation is unveiled, for a few selected values
of the coupling constants. We succeed in identifying critical points
and deriving local order parameters to characterize different phases
in the conventional Ginzburg-Landau-Wilson paradigm.
\end{abstract}


\maketitle
{\it Introduction.} Quantum phase transitions
(QPTs)~\cite{sachdev,wen}, occurring at absolute zero temperature,
are driven by quantum fluctuations in quantum many-body systems. In
condensed matter, a theoretical description of such a system usually
involves two-body interactions. However, recent progress makes it
possible to realize multi-body interactions  with current technology
in experiments, which, as theoretical analyses unveiled, produce
exotic quantum phases. Examples include a spin model with competing
two-spin and three-spin interactions in a system of trapped
ions~\cite{aberm} and of ultracold atoms in triangular
lattices~\cite{jkmb} and cold polar molecules~\cite{hpbam}.
Unfortunately, the presence of multi-body interactions in a spin
model normally renders it not exactly solvable. This limits the
applicability of conventional analytic approaches. Therefore, it is
desirable to develop approximated techniques or fully numerical
approaches to address such a novel many-body system.

In the past few years, significant progress has been made to develop
efficient numerical algorithms to simulate quantum many-body lattice
systems in the context of the tensor network
representations~\cite{vidal1,vidal2}, resulting from advances in our
understanding of quantum entanglement present in ground-state wave
functions for quantum many-body lattice systems~\cite{amico}. In
Ref.~\cite{vidal1}, Vidal introduced the infinite time-evolving
block decimation (iTEBD) algorithm to find an approximate ground
state for an infinite-size one-dimensional (1D) lattice system in
the infinite matrix product state (iMPS) representation. In
Ref.~\cite{vidal2}, an infinite projected entangled-pair state
(iPEPS) is proposed for an infinite-size quantum system in two and
higher spatial dimensions. A peculiar feature of these algorithms is
their ability to compute the ground-state fidelity per lattice
site~\cite{zhou}-~\cite{ywd1}. Fidelity describes the distance
between two given quantum states, thus enabling us to capture
drastic changes in ground-state wave functions for quantum many-body
lattice systems undergoing QPTs~\cite{zhou}-~\cite{fidelity0}. In
Refs.~\cite{zhou,zov}, it was argued that the ground-state fidelity
per lattice site is able to characterize QPTs, regardless of what
type of internal order is present in quantum many-body states
underlying various quantum lattice systems in condensed matter.
However, an adaptation of the algorithms is necessary to make it
possible to investigate a spin model with multi-body interactions.

In this work, we first adapt the iTEBD algorithm, making it suitable
to generate a ground-state wave function for a 1D
translation-invariant spin model with the competing two-body and
three-body interactions in the context of the iMPS representation.
Then the ground-state fidelity per lattice site is computed, and its
bifurcations is unveiled for different values of the truncation
dimension, with a bifurcation point as a pseudo-critical point,
which tends to the critical point when the truncation is removed. In
addition, the bifurcation implies symmetry spontaneous breaking,
characterized by a local order parameter, a central concept in the
conventional Landau-Ginzburg-Wilson paradigm~\cite{pwand,scole}. The
latter is read off from a reduced density matrix of a representative
ground-state wave function from a given phase.

\begin{figure}
\begin{center}
\includegraphics[width=0.48\textwidth]{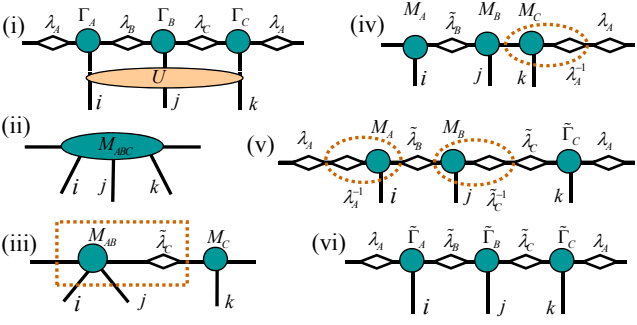}
\caption{(color online) The updating steps of an infinite matrix
product state (iMPS) algorithm adapted to a spin model with
three-body interactions. Here, $\Gamma_{\eta}^{s}$ and
$\lambda_{\eta}$ ($\eta=A,B,C$) are three-index tensors and diagonal
matrices, respectively. (i) Apply a three-site gate $U_{i,j,k}$ to
the cell. (ii) Contract all the tensors in (i) and reshape the
tensor $M_{ABC}$ into a matrix. (iii) Perform the singular value
decomposition (SVD) to yield  $M_{AB}$, $\tilde{\lambda_{C}}$, and
$M_{C}$. (iii)  Contract the tensor $M_{AB}$ and
$\tilde{\lambda_{C}}$ and reshape it into a matrix. (iv) Perform the
SVD to yield $M_{A}$, $\tilde{\lambda_{B}}$, and $M_{B}$. (v)
Restore the iMPS representation with the identity resolution
$\lambda_{\eta} \lambda_{\eta}^{-1}= {\rm id}$ inserted. (vi) The
diagonal matrices $\tilde{\lambda_{B}}$, $\tilde{\lambda_{C}}$, and
tensors $\tilde{\Gamma_{A}}$, $\tilde{\Gamma_{B}}$, and
$\tilde{\Gamma_{C}}$ are updated. as indicated by the dash-line
ovals in (iv) and (v).} \label{Fig1}
\end{center}
\end{figure}

{\it Infinite matrix product state algorithm adapted to a spin model
with three-body interactions.} We first describe an iMPS algorithm
adapted to a spin model with three-body interactions. For an
infinite-size translation-invariant quantum many-body system with
three-body interactions described by a Hamiltonian
$H=\sum_{i}h^{[i,i+1,i+2]}$, we introduce the iMPS representation
that is translation-invariant under a three-site shift. That means
we need three three-index tensors $\Gamma_{Alr}^{i}$,
$\Gamma_{Blr}^{j}$, and $\Gamma_{Clr}^{k}$ and three diagonal
matrices $\lambda_{A}$, $\lambda_{B}$, and $\lambda_{C}$  to store a
wave function in the context of the iMPS representation, as shown in
Fig.~\ref{Fig1}(i). Here, $i,j$, and $k$ are physical indices, which
run over a $d$-dimensional local Hilbert space, $l$ and $r$ are the
inner bond indices, $l$, $r$ =1,..., $\chi$,  with $\chi$ being the
truncation dimension. This amounts to the statement that the unit
cell of the iMPS representation consists of three consecutive sites.
The ground-state wave function is projected out by resorting to the
imaginary time evolution.

The updating procedure of the algorithm is as follows. Apply a
three-site imaginary time evolution gate
$U(i,i+1,i+2)=\exp(-h^{[i,i+1,i+2]\Delta\tau})$ over a time slice,
$\Delta\tau\ll1$, to the unit cell of the iMPS representation, and
contract all the tensors, as shown in Fig.~\ref{Fig1}(i). As such,
we have a tensor $M_{ABC}$ (Fig.~\ref{Fig1} (ii)). Reshape the
tensor $M_{ABC}$ into a $\chi d^{2}\times\chi d$ matrix and perform
the singular value decomposition (SVD) to the matrix, we get
$M_{AB}$, $\tilde{\lambda_{C}}$, and $M_{C}$, as shown in
Fig.~\ref{Fig1} (iii). Contract the tensor $M_{AB}$ and
$\tilde{\lambda}_{C}$ and reshape it into a $\chi d\times\chi d$
matrix and perform the SVD, we have $M_{A}$, $\tilde{\lambda}_{B}$,
and $M_{B}$, as seen in Fig.~\ref{Fig1} (iv). Insert the identity
resolution $\lambda_A \lambda_A^{-1}= {\rm id}$ and contract $M_{C}$
and $\lambda_{A}^{-1}$ to yield $\tilde{\Gamma}_{C}$. In
Fig.~\ref{Fig1} (v), the identity resolution $\lambda_{\eta}
\lambda_{\eta}^{-1}= {\rm id}$ ($\eta=A$ and $C$) is inserted, thus
we are able to update the diagonal matrices $\tilde{\lambda_{B}}$,
$\tilde{\lambda_{C}}$, and tensors $\tilde{\Gamma_{A}}$,
$\tilde{\Gamma_{B}}$, and $\tilde{\Gamma_{C}}$. Once this is done,
we shift one site, apply the gate $U(i+1,i+2,i+3)$, and repeat the
updating procedure. Afterwards, shift one site, apply the gate
$U(i+2,i+3,i+4)$ and repeat the updating procedure again. This
completes our updating procedure for a time slice $\Delta\tau$.
Iterating until the ground-state energy is converged, the system's
ground-state wave function is yielded in the iMPS representation.
\begin{figure}
\begin{center}
\includegraphics[width=0.33\textwidth]{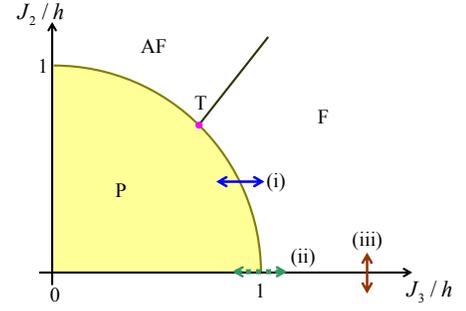}
\caption{(color online) Schematic phase diagram for a 1D spin model
with the competing two-spin and three-spin interactions. For $J_2>0$
and $J_3>0$, there are three phases: a disordered paramagnetic (P)
regime, an antiferromagnetic (AF) order phase, and  a ferrimagnetic
(F) phase. Three phases coexist in the tricritical point T. In
addition, the model becomes quantum Ising chain in a transverse
field if $J_3=0$. We focus on three typical cases: (i) The
horizontal double arrow crosses the phase boundary from P to F along
$J_{2}/h$=0.4. (ii) The horizontal dotted double arrow crosses a
critical point along $J_{2}/h$=0. (iii) The vertical double arrow
crosses a discontinuous QPT point along $J_{3}/h$=1.5.}\label{Figa}
\end{center}
\end{figure}

{\it Model.} We consider a 1D spin model with the competing two-spin
and three-spin interactions~\cite{aberm}, which is described by the
Hamiltonian:
\begin{equation}
  H=J_{2}\sum_{i}\sigma^{[i]}_{z}\sigma^{[i+1]}_{z}+J_{3}\sum_{i}
  \sigma^{[i]}_{z}\sigma^{[i+1]}_{z}\sigma^{[i+2]}_{z}-
  h\sum_{i}\sigma^{[i]}_{x},\label{ham2}
\end{equation}
where $J_{2}$ and $J_{3}$ are, respectively, the two-spin and
three-spin coupling constants, $h$ is an external field along the
$x$ direction, and $\sigma^{[i]}_{\alpha} \:(\alpha=x,z)$ are the
spin $1/2$ Pauli operators at the $i$-th site. If the coupling
constants are varied independently, it undergoes QPTs, with the
occurrence of different quantum phases: an antiferromagnetic (AF)
phase, a ferrimagnetic (F) phase, and a disordered paramagnetic (P)
regime where spins are aligned along the $x$ direction~\cite{aberm}.
Note that, if $J_{3}=0$, the model Hamiltonian $H$ becomes quantum
Ising model in a transverse field, which is exactly solvable, with
the critical point at $h/J_{2}=1$. A schematic phase
diagram~\cite{aberm} is shown in Fig.~\ref{Figa}. Here, we fix $h$=1
and focus on three cases: (i) $J_{2}$=0.4, $J_{3}$ varies from
$J_{3}=0.80$ to $J_{3}=1.00$ (horizontal solid double arrow in
Fig.~\ref{Figa}). (ii) $J_{2}$=0, $J_{3}$ varies from $J_{3}=0.90$
to $J_{3}=1.10$ (horizontal dotted double arrow  in
Fig.~\ref{Figa}). (iii) $J_{3}$=1.5, $J_{2}$ varies from
$J_{2}=-0.1$ to $J_{2}=0.1$ (vertical solid double arrow  in
Fig.~\ref{Figa}).
\begin{figure}
\begin{center}
\includegraphics[width=0.54\textwidth]{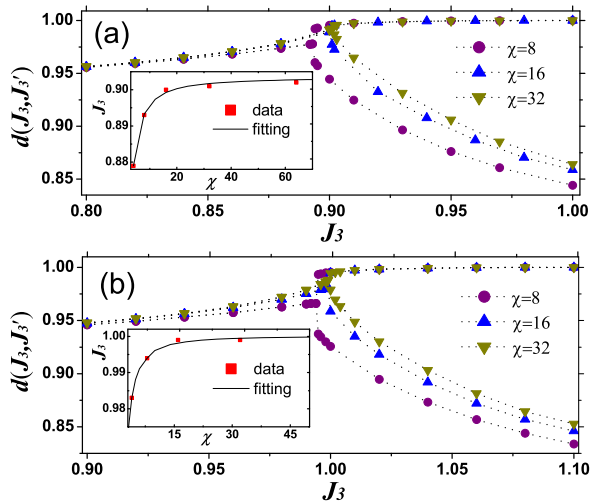}
\caption{(color online) Main: The ground-state fidelity per lattice
site, $d(J_{3},J_{3}')$, as a function of $J_{3}$, at a fixed
$J_{3}'$, for a 1D spin model with the competing two-spin and
three-spin interactions. Here, $J_{2}=0.4$ (a) and $J_{2}=0$ (b),
respectively, and $J_{3}$ is chosen as the control parameter. In
addition, $|\varphi(J_{3}')\rangle$, with $J_{3}'=1.00$ and
$J_{3}'=1.10$, is chosen as a reference state, respectively, in (a)
and (b). We observe that there is a bifurcation point in the
ground-state fidelity per lattice site, indicating that spontaneous
symmetry breaking occurs when the control parameter $J_{3}$ varies.
The bifurcation point is a pseudo critical point for a given value
of the bond dimension. Inset: An extrapolation with respect to the
bond dimension is performed for the pseudo critical points for
$\chi$=4, 8, 16, 32, and 64, yielding the critical points $J_{3c}
\sim 0.9034$ (a) and $J_{3} \sim 1.0001$ (b), when
$\chi\rightarrow\infty$. } \label{Fig2}
\end{center}
\end{figure}

\begin{figure}
\begin{center}
\includegraphics[width=0.46\textwidth]{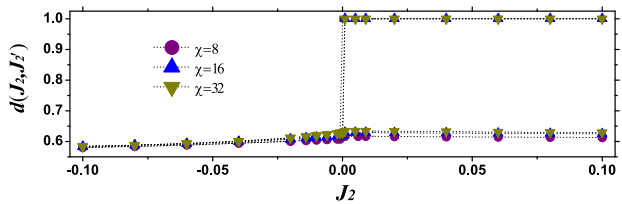}
\caption{(color online) The ground-state fidelity per lattice site,
$d(J_{2},J_{2}')$, as a function of $J_{2}$, at a fixed $J_{2}'$,
for a 1D spin model with the competing two-spin and three-spin
interactions. Here, $J_{3}=1.5$, and $J_{2}$ is chosen as the
control parameter. In addition, the reference state
$|\varphi(J_{2}')\rangle$ is fixed at $J_{2}'=0.10$. Spontaneous
symmetry breaking occurs with $J_{2}>0$. A jump occurs in the
ground-state fidelity per lattice site, meaning that the phase
transition is discontinuous. The pseudo critical points are shown
for different values of the truncation dimension, without any
significant shift from $\chi=8$ to $\chi=32$.} \label{Fig4}
\end{center}
\end{figure}

{\it Bifurcation in the ground-state fidelity per lattice site and
symmetry spontaneous breaking.} For two ground-state wave functions
$|\varphi(\lambda_{1})\rangle$ and $|\varphi(\lambda_{2})\rangle$,
with $\lambda_{1}$ and $\lambda_{2}$ being two values of the control
parameter $\lambda$, the ground-state fidelity
$F(\lambda_{1},\lambda_{2})=|\langle\varphi(\lambda_{1})|\varphi(\lambda_{2})\rangle|$
asymptotically scales as $F(\lambda_{1},\lambda_{2})\sim
d(\lambda_{1},\lambda_{2})^{L}$, where $L$ is the number of the
lattice sites. The scaling parameter, $d(\lambda_{1},\lambda_{2})$,
first introduced in Ref.~\cite{zhou}, characterizes how fast the
ground-state fidelity goes to zero when the thermodynamic limit is
approached. Physically, $d(\lambda_{1},\lambda_{2})$ is the {\it
averaged} ground-state fidelity per lattice site,
\begin{equation}
\ln d(\lambda_{1},\lambda_{2}) = \lim_{L
\rightarrow\infty}\frac{F(\lambda_{1},\lambda_{2})}{L}.
\end{equation}
As discussed in Refs.~\cite{zhou,zov}, it is well defined in the
thermodynamic limit, although $F(\lambda_{1},\lambda_{2})$ becomes
trivially zero. The ground-state fidelity per lattice site is able
to locate a QPT in a quantum many-body lattice system, regardless of
what type of internal order is present.

The iMPS representation allows to efficiently compute the
ground-state fidelity per lattice site~\cite{zov}. Now we turn to
the simulation results for the three cases, as selected above, of
the model Hamiltonian (\ref{ham2}) in the context of the iMPS
algorithm.

(i) $J_{2}=0.4$ and $h=1$, and $J_{3}$ is chosen as the control
parameter. The ground-state fidelity per lattice site is plotted in
Fig.~\ref{Fig2}(a), with the truncation dimension $\chi=$8, 16, and
32, respectively. Note that, $|\varphi(J_{3}')\rangle$, with
$J_{3}'=1.00$, is chosen as a reference state. We observe that there
is a bifurcation point in the ground-state fidelity per lattice
site, arising from  spontaneous symmetry breaking  of the $Z_3$
translational invariance under three-site shifts, when the control
parameter $J_{3}$ varies from $J_{3}=0.80$ to $J_{3}=1.00$. The
bifurcation point is a pseudo critical point for a given value of
the truncation dimension. The extrapolation of the pseudo critical
points for $\chi$=4, 8, 16, 32, and 64 yields the critical point
$J_{3c}=0.9034$, when $\chi\rightarrow\infty$. Therefore, a
continuous QPT occurs at $J_{3c} \sim 0.9034$.

(ii) $J_{2}=0$ and $h=1$, and $J_{3}$ is chosen as the control
parameter. The ground-state fidelity per lattice site,
$d(J_{3},J_{3}')$, as a function of $J_{3}$, at a fixed $J_{3}'$, is
plotted in  Fig.~\ref{Fig2}(b), with the truncation dimension
$\chi=$ 8, 16, and 32, respectively.  The reference state
$|\varphi(J_{3}')\rangle$ is chosen at $J_{3}'=1.10$. We observe
that there is a bifurcation point in the ground-state fidelity per
lattice site. Actually, four degenerate ground states arise from
spontaneous symmetry breaking of the group $G=Z_2 \times Z_2 \times
Z_2/Z_2$, when the control parameter $J_{3}$ varies from
$J_{3}=0.90$ to $J_{3}=1.10$. Here, four $Z_2$ groups are,
respectively, generated from $\sigma_{z}^{A}\sigma_{z}^{B}$,
$\sigma_{z}^{B}\sigma_{z}^{C}$, $\sigma_{z}^{A}\sigma_{z}^{C}$, and
$\sigma_{z}^{A}\sigma_{z}^{B}\sigma_{z}^{C}$, where $A, B$, and $C$
label three sites in the unit cell, as shown in Fig.~\ref{Fig1}.
Note that a pseudo critical point $J_{3\chi}$ is identified as a
bifurcation point. Therefore, a continuous QPT occurs at $J_{3} \sim
1.0001$.

(iii) The ground-state fidelity per lattice site, $d(J_{2},J_{2}')$,
as a function of $J_{2}$, at a fixed $J_{2}'$, is plotted in
Fig.~\ref{Fig4} for $J_{3}=1.5$, with $J_{2}$ as the control
parameter. In addition, the reference state
$|\varphi(J_{2}')\rangle$ is fixed at $J_{2}'=0.1$. Spontaneous
symmetry breaking of the translation group $Z_3$ under three-site
shifts occurs for $J_{2}>0$. There is a jump at $J_2=0$ in the
ground-state fidelity per lattice site, meaning that the phase
transition is discontinuous. The pseudo critical points are shown
for different values of the truncation dimension, without any
significant shift, when $\chi$ is increased from $\chi=8$ to
$\chi=32$.

\begin{figure}
\begin{center}
\includegraphics[width=0.54\textwidth]{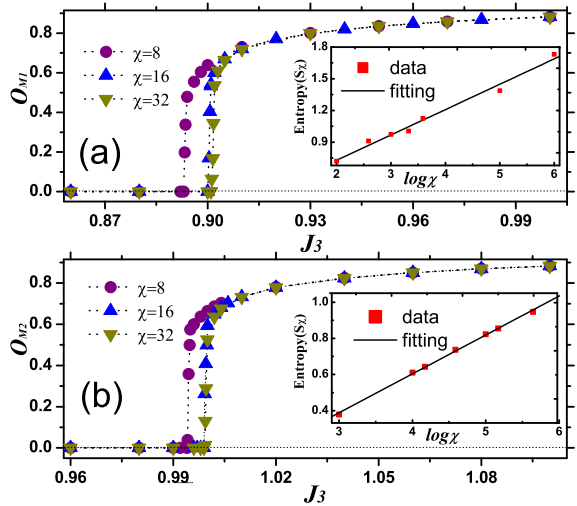}
\caption{(color online) Main: The local order parameter
$\textbf{\textit{O}}_{M1}=(\langle\sigma_{z}^{A}\rangle-2\langle\sigma_{z}^{B}\rangle+\langle\sigma_{z}^{C}\rangle)/4$
shown in (a) and $\textbf{\textit{O}}_{M2}=\langle\sigma_{z}\rangle$
shown in (b) for a 1D spin model with the competing two-spin and
three-spin interactions, where $J_{2}=0.4$ and $J_{2}=0$,
respectively. The computation is performed with the truncation
dimension $\chi=8, 16$ and 32. The pseudo critical points located by
the order parameter are consistent with those obtained by the
ground-state fidelity per lattice site. We see that
$\textbf{\textit{O}}_{M1}$ and $\textbf{\textit{O}}_{M2}$  tend to
saturate, with increasing $\chi$. Inset: The extrapolation of the
semi-infinite chain von Neumann entropy  is performed, yielding the
central charge $c \sim 0.796$, with respect to the bond dimension
$\chi=4, 6, 8, 10, 12, 32$ and $64$, in (a) and $c \sim 0.994$, with
respect to the bond dimension $\chi= 8, 16, 18, 24, 32, 36$ and
$50$, in (b), very close to the exact results $c=4/5$ and $1$,
respectively.} \label{Fig5}
\end{center}
\end{figure}

{\it Local order parameters.} In the conventional
Ginzburg-Landau-Wilson paradigm, a central concept is local order
parameters, whose nonzero values characterize symmetry broken
phases. As advocated in Ref.~\cite{zhou3}, any local order parameter
may be derived from a reduced density matrix from a representative
ground-state wave function in a symmetry broken phase, if its iMPS
representation is known. Actually, this follows from the
non-zero-entry structure of a reduced density matrix: in a symmetric
phase, any reduced density matrix respects all the symmetry of the
system, but in a symmetry broken phase, there are extra nontrivial
entries in a reduced density matrix due to the fact that a symmetry
operation is lost. Following this line of reasoning, we are able to
derive local order parameters for different symmetry broken phases.

In the $Z_3$ symmetry broken phase, three degenerate ground states
arise.  The non-zero-entry structure of the one-site reduced density
matrix yields a local order parameter
$\textbf{\textit{O}}_{M1}=(\langle\sigma_{z}^{A}\rangle-2\langle\sigma_{z}^{B}\rangle+\langle\sigma_{z}^{C}\rangle)/4$,
which is shown in Fig.~\ref{Fig5}(a). In fact, we may equally choose
any of the following three operators as a local order parameter:
$(-2\langle\sigma_{z}^{A}\rangle+\langle\sigma_{z}^{B}\rangle+\langle\sigma_{z}^{C}\rangle)/4$,
$(\langle\sigma_{z}^{A}\rangle-2\langle\sigma_{z}^{B}\rangle+\langle\sigma_{z}^{C}\rangle)/4$,
and
$(\langle\sigma_{z}^{A}\rangle+\langle\sigma_{z}^{B}\rangle-2\langle\sigma_{z}^{C}\rangle)/4$,
where $A, B$, and $C$ label three sites in the unit cell, as shown
in Fig.~\ref{Fig1}.

In the $G$ symmetry broken phase, four degenerate ground states
arise. A local order parameter may be constructed from an analysis
of the non-zero-entry structure of the one-site reduced density
matrix. One may choose $\langle\sigma^A_{z}\rangle$,
$\langle\sigma^B_{z}\rangle$, and $\langle\sigma^C_{z}\rangle$ as a
local order parameter, which are able to distinguish four degenerate
ground states. Their magnitudes take the same value, which is shown
in Fig.~\ref{Fig5}(b).

Now we take advantage of the finite-entanglement scaling of the von
Neumann entropy to determine the nature of the criticalities.  The
von Neumann entropy $S$ for a semi-infinite chain scales as $S \sim
c \kappa/6 \log \chi$, where $c$ is the central charge, and $\kappa$
is the finite-entanglement scaling exponent~\cite{pp}.  Our best
fitting results, as shown in the insets of Fig.~\ref{Fig5} (a) and
(b), yields the central charge $c \sim 4/5$ and $c \sim 1$, with
$\kappa \sim 1.8077 $ and $\kappa \sim 1.2996$, respectively. This
implies that the transition points $J_{3c} \sim 0.9034$ and $J_{3c}
\sim 1.0001$ are, respectively, in the same universality class as
the three-state Potts universality class and the four-state Potts
universality class.

{\it Conclusions.} We have investigated a quantum spin model with
the competing two-spin and three-spin interactions by exploiting a
tensor network algorithm based on the iMPS representation. The
algorithm itself is an adaptation of Vidal's iTEBD algorithm to a
translation-invariant 1D lattice spin system with three-spin
interactions. This enables us to efficiently compute the
ground-state fidelity per lattice site, which in turn makes it
possible to locate critical points by identifying bifurcation points
in the ground-state fidelity per lattice site. For a few selected
values of the coupling constants, we succeeded in identifying
critical points, and deriving local order parameters to characterize
different phases in the conventional Ginzburg-Landau-Wilson
paradigm.

{\it Acknowledgements.} We thank Sam Young Cho, Bo Li, Sheng-Hao Li,
Qian-Qian Shi, Ai-Min Chen, Yao-Heng Su, Jian-Hua Liu and Jian-Hui
Zhao for helpful discussions. The work is partially supported by the
National Natural Science Foundation of China (Grant No: 10874252).
HLW and YWD are supported by the Fundamental Research Funds for the
Central Universities (Project No. CDJXS11102214) and by Chongqing
University Postgraduates' Science and Innovation Fund (Project No.:
200911C1A0060322).

\bibliographystyle{elsarticle-num}
\bibliography{<your-bib-database>}

\end{document}